\documentstyle[aps,epsf,twocolumn,header]{revtex}

\begin{document}
\draft
\title
      { Spinning Down A Black Hole With Scalar Fields}
\author
      { Chris M. Chambers
         \thanks{Electronic Address: chrisc@orion.physics.montana.edu},
        William A. Hiscock
	 \thanks{Electronic Address: billh@orion.physics.montana.edu}
	and
	Brett Taylor
         \thanks{Electronic Address: brett@peloton.physics.montana.edu}}
\address
      {Department Of Physics,
       Montana State University,
       Bozeman,
       Montana. MT 59717-3840}
\date
      {Received \today}
\preprint{MSUPHY97.0xx}
\maketitle

%
%

\begin{abstract}
We study the evolution of a Kerr black hole emitting
scalar radiation via the Hawking process. We show that
the rate at which mass and angular momentum are lost
by the black hole leads to a final evolutionary state with
nonzero angular momentum, namely $a/M \approx 0.555$.
\end{abstract}

\pacs{PACS number(s): 04.62.+v,04.25.Dm,04.70.-s,04.70.Dy}

\tighten

%
%

\section{Introduction}

The conventional view of black hole evaporation is that, regardless
of its initial state, Hawking radiation will cause a black hole to
approach an uncharged, zero angular momentum state long before
all its mass has been lost. Thus, as the evolution nears the Planck
scale, where quantum gravity will be required to determine its future
development, the final asymptotic state is assumed to be described by the
Schwarzschild solution.

Although calculations have shown that any charge on the black hole will
be rapidly neutralized~\cite{Gibbons}, qualitative arguments~\cite{carter1:74}
suggest that the mass and angular momentum
loss proceeds on cosmologically long time scales. Such arguments, however,
do not determine if the angular momentum will tend to 
zero before the mass does. Using Hawking's results on quantum
emission from black holes\cite{hawking1:75}, Page has been able to 
address this and other related questions. 
By performing an exhaustive, quantitative, study
of the emission of particles associated with fields of spin $1/2,1$
and $2$ for both rotating \cite{page1:76} and non-rotating
holes \cite{page2:76}, he concludes that for these fields the
specific angular momentum does tend to zero more rapidly than the mass. 
This implies  that the final asymptotic state of black
hole evaporation is indeed the Schwarzschild solution. 

However, Page did provide some indirect evidence to suggest that if 
there were a sufficiently large number of, as yet unknown, massless 
scalar fields present in nature, then the dimensionless ratio of the 
black hole's specific angular momentum to its mass, $a_{*} =a/M$, 
might evolve to a stable nonzero value. In this case, a microscopic 
evaporating black hole approaching the Planck scale would be described
by a Kerr solution, characterized by a mass $M$ and specific angular 
momentum $a$, rather than Schwarzschild. Firstly, at low 
$a_{*}=a/M$, Page found that the dominant angular 
modes for the spin-$s$ fields, $s=1/2,1$ and $2$, were those with $l=s$. 
If the same is true for the scalar field, with $s=0$, then the 
dominant mode would be $l=0$, which carries off energy, but no
angular momentum. In that case, 
one could imagine a scenario in which the mass loss occurs much more
rapidly than the angular momentum loss, causing the black hole 
to evolve toward a state with $a_{*}$ nonzero. Secondly, Page noted
in his results a simple, perhaps accidental linear relationship 
between the spin of the field and the (suitably normalized) ratio of
the angular momentum loss rate to the mass loss rate, $h(a_{*})$,  
as $a_{*}$ tended to zero. Page pointed out that if this
relation were extrapolated to the case of a scalar field, $s=0$, 
then the loss rates for emission of scalar particles would result
in a final state of nonzero angular momentum. 

In this letter we investigate, in some detail, the evolution of 
a Kerr black hole emitting scalar radiation via
the Hawking process. For clarity we have restricted our attention
to the mass and angular momentum loss rates, and their effect on
the final evolutionary state of the hole. We find that the hole
does indeed evolve to a final asymptotic 
state with nonzero angular momentum,
confirming the conjecture of Page. Our numerical results 
allow us to conclude that the final state will be 
described by a specific angular momentum $a = 0.555 M$. 
In addition, we find that the linear relationship between the spin
of the field and $h(a_{*})$, found by Page, breaks down for $s=0$. 
We find that for a scalar field, $h(a_{*}=0)=-0.806$, rather than the value 
$-1.195$ which Page obtained by extrapolation. The details of this 
work will appear elsewhere.

The notation here and throughout follows that of Press and
Teukolsky~\cite{press1:74} and Page~\cite{page1:76}. We make
the assumption that the black hole has existed for a 
sufficient period of time so that any charge it possessed has been
lost, and hence is adequately described by a Kerr solution.
In terms of Kerr-ingoing coordinates $(v,r,\theta,\tilde{\phi})$, 
the scalar wave equation, $\Box \phi=0$, separates by writing 
$\phi=R(r) S(\theta) e^{-i \omega v} e^{i m \tilde{\phi}}$, where
the angular function $S(\theta)$ is a spheroidal harmonic~\cite{flammer}.
The radial function, $R(r)$, satisfies
	\begin{equation}
	  ( \partial_{r} \Delta \partial_{r} - 2 i K
	    \partial_{r}
	    - 2 i \omega r  -\lambda ) 
	    R(r) = 0 \label{1.2} \; ,
	\end{equation}
where $\Delta = r^2 -2Mr+a^2$, $K=(r^2+a^2) \omega - am$, $\lambda=
E_{lm\omega}-2am\omega+a^2 \omega^2$  and $E_{lm\omega}$
is the separation constant.
While the solutions to Eq.~(\ref{1.2}) are not, in general, expressible in 
terms of known functions, their asymptotic behavior is easily 
obtained \cite{press1:74},
	\begin{equation}
	  R \longrightarrow \left\{ \begin{array}{ll}
	    Z_{{\rm hole}}  &  r \rightarrow r_{+} \\
	    Z_{{\rm in}} r^{-1} + Z_{{\rm out}} r^{-1}
	    e^{2 i \omega r} & r \rightarrow \infty
				    \end{array}
			    \right. \label{1.3} \; .
	\end{equation}
The subscript `in' refers to an ingoing wave originating from past null
infinity, `out' refers to the reflected component of the 
wave that propagates outward, toward future null infinity, and `hole'
refers to the transmitted component that crosses the black hole
event horizon at $r=r_{+}$.  The amplification, $Z$ 
(the fractional gain of energy in a scattered wave), is
	\begin{equation}
	  Z= \left| \frac{Z_{\rm out}}{Z_{\rm in}} \right| -1
	     \label{1.4} \; .
	\end{equation}
Page \cite{page1:76} has shown that, in terms of the scale 
invariant quantities $f \equiv -M^2 dM/dt$ and $g \equiv -M a^{-1}_{*}
dJ/dt$, the rate at which the mass and angular momentum of
an evaporating black hole decrease is given by
	\begin{equation}
	  \left( \begin{array}{c}
		   f \\ g
		 \end{array}
	  \right) =
	    -\sum_{l,m} \frac{1}{2\pi} \int^{\infty}_{0}
	    dx \frac{Z}{e^{2 \pi k / \kappa} -1}
	  \left( \begin{array}{c}
		    x \\ m a_{*}^{-1} 
		 \end{array}
	  \right) \label{1.5} \; ,
	\end{equation}
where  $k=\omega-m \Omega$, $\Omega=a_{*}/2r_{+}$ is the surface
angular frequency, $\kappa=\sqrt{(1-a_{*}^{2})}/2r_{+}$ is the
surface gravity of the hole, and following Page \cite{page1:76}
we have defined $x=M \omega$ and $a_{*}=a/M$. In order to 
investigate how $a_{*}$ varies with $M$ as the black hole
loses mass and angular momentum, we define the function $h(a_{*})$
	\begin{equation}
	  h(a_{*}) \equiv \frac{ d \ln a_{*} }{ d \ln M }
	  = \frac{g(a_{*})}{f(a_{*})} - 2  \label{1.6} \; .
	\end{equation}
The rate of change of $a_{*}$ is then given by
	\begin{equation}
	{da_{*} \over dt} = -{a_{*}fh \over M^3} .
	\label {1.7}
	\end{equation}
If there is a nonzero value of $a_{*}$ for which $h$ is zero, then
$da_{*}/dt$ will be zero there. Since $f$ is nonnegative, if
$h$ is positive above this value and negative below it, then
an evaporating black hole will evolve towards a stable state
at $h(a_{*})=0$.
We now have all the necessary machinery needed to study the 
mass and angular momentum loss rates of a purely rotating
black hole.

For clarity we have limited our attention to the behavior of
the mass and angular momentum loss rates, and their effect
on the final asymptotic state of the evaporating hole. We calculated 
the functions $f(a_{*})$ and $g(a_{*})$ at 18 values of $a_{*}$ 
ranging from $a_{*}=1 \times 10^{-4}$ to $a_{*}=0.99$ and used a
clamped cubic spline to extrapolate these values to $a_{*}=0$
and $a_{*}=1$. The same spline was used to interpolate for
points of interest.

%
%
\begin{figure}
\leavevmode
\begin{center}
\begin{minipage}{0.5\textwidth}
\epsfxsize=0.95\textwidth \epsfysize=0.95\textwidth {\epsffile{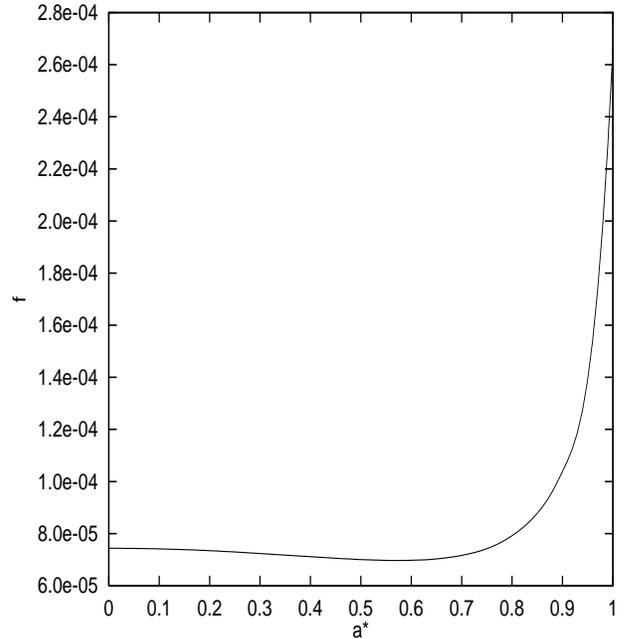}}
\end{minipage}
\end{center}
\vspace{0.5cm}
\caption{The mass loss rate, due to scalar particle emission, for
a Kerr black hole. At low rotation the mass loss rate
approaches the value $7.439 \times 10^{-5}$, while at
the extreme limit $a_{*}=1$, it reaches $2.601 \times 10^{-4}$.
There is a minimum in the emission at $a_{*} = 0.574$.}
\label{fig:1}
\end{figure}
%
%

Figure~\ref{fig:1} shows the behavior of the mass loss rate 
as a function of the specific angular momentum, described in 
a scale invariant way by the function $f(a_{*})$.
The loss of mass-energy from the hole by emission of scalar
particles is more effective at high values of $a_{*}$. The fact 
that emission still occurs at $a_{*}=1$, even
though the hole has zero temperature, is due to the nonzero
chemical potential associated with the angular momentum of
the hole; this results in spontaneous emission into the
superradiant modes, first discovered by Zel'dovich~\cite{zeldo:71}.
An interesting feature of Fig.~\ref{fig:1} is the existence of 
a minimum at $a_{*}=0.574$, which does not occur for fields
of nonzero spin. The unusual behaviour of the mass loss, in
this case, is mainly attributable to the fact that a scalar
field, unlike higher spin fields, is able to radiate in an
$l=0$ mode, which is not a superradiant mode.
A plot of the mass loss rate, due solely to the $l = 0$ mode
of the field, reveals $f_{l = 0}$ to be a monotonically
decreasing function of $a_{*}$. This suggests that the hole
couples most strongly to the $l=0$ mode at low rotation. At
larger values of $a_{*}$ the contribution to $f$ from the
higher $l$-modes becomes more significant, as the effects of
superradiant scattering increase. This has the effect of 
increasing the mass loss rate, hence $f$, at high $a_{*}$. 
The combination of these two effects, at intermediate $a_{*}$, 
results in the appearance of a minimum in $f$ as shown. 
%
%
\begin{figure}
\leavevmode
\begin{center}
\begin{minipage}{0.5\textwidth}
\epsfxsize=0.95\textwidth \epsfysize=0.95\textwidth {\epsffile{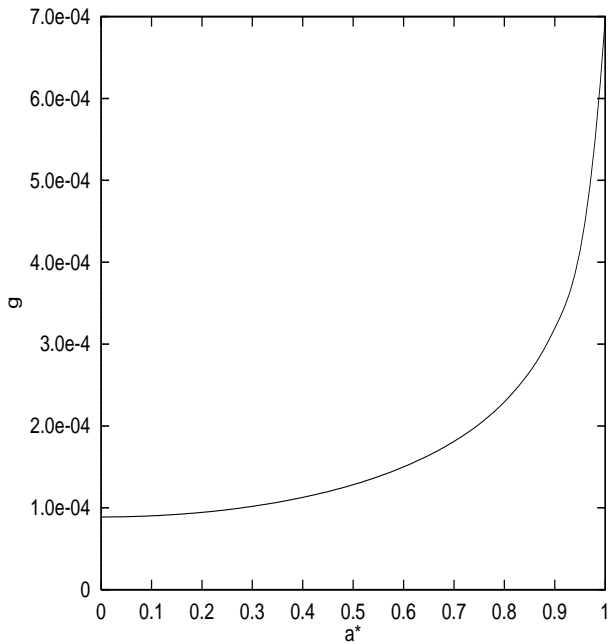}}
\end{minipage}
\end{center}
\vspace{0.5cm}
\caption{The angular momentum loss rate, due to scalar particle
emission, for a Kerr black hole. At low rotation, the angular
momentum loss rate approaches $8.886 \times 10^{-5}$. At the
extreme limit, $a_{*}=1$, the emission rate is $6.853 \times
10^{-4}$.}
\label{fig:2}
\end{figure}
%
%

Figure~\ref{fig:2} displays
the angular momentum loss rate, $g$, versus $a_{*}$. Since the 
$l=0$ mode cannot carry off any angular momentum, the behavior
of $g$ can be understood purely in terms of superradiance. 
As $a_{*}$ increases emission into the superradiant modes becomes 
more effective, causing the angular momentum loss rate to 
increase monotonically. Again, at the extreme limit, scalar
particle emission continues to carry off angular momentum despite
the temperature of the hole being zero, by spontaneous emission 
into the superradiant modes.
%
%
\begin{figure}
\leavevmode
\begin{center}
\begin{minipage}{0.5\textwidth}
\epsfxsize=0.95\textwidth \epsfysize=0.95\textwidth {\epsffile{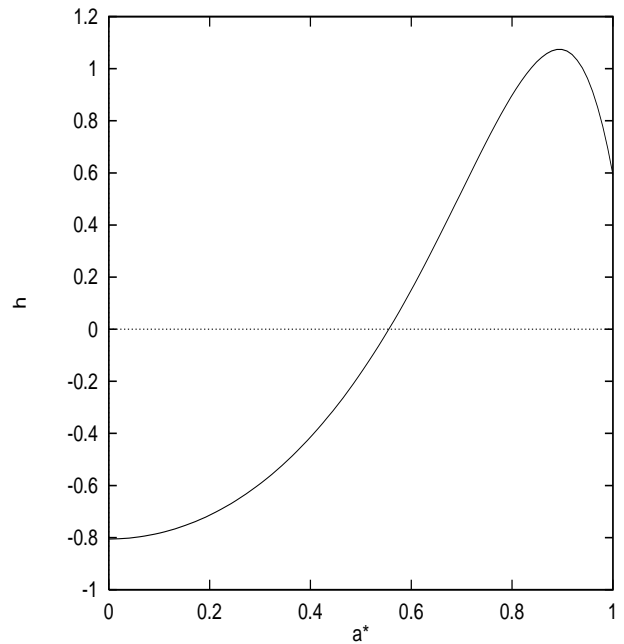}}
\end{minipage}
\end{center}
\vspace{0.5cm}
\caption{The angular
momentum loss rate compared to the mass loss rate,
described in a scale invariant way by $h(a_{*})$.
The point at which $h(a_{*}) = 0$ occurs at
$a_{*}=0.555$. A hole formed with $a_{*}$ on either
side of this value will evolve to a state characterized
by this value. As $a_{*} \rightarrow 0$, $h(a_{*})$
approaches the value $-0.806$}
\label{fig:3}
\end{figure}
%
%

Finally, Figure~\ref{fig:3} shows the behavior of $h(a_{*})$. The most 
important feature is the existence of a zero in $h$, at $a_{*}=0.555$. 
A black hole formed with $a_{*} < 0.555$ or $a_{*}>0.555$
will evolve, by  the loss of mass and angular momentum, 
until it reaches an asymptotic state 
characterised by $a_{*}=0.555$. This confirms the conjecture made 
by Page~\cite{page1:76}, that a black hole emitting scalar radiation, 
via the Hawking process, could evolve to a final state with nonzero 
angular momentum.

From his examination of spin $1/2$, $1$ and $2$ fields, Page found
a remarkable, linear relationship between $h(a_{*})$ and the
spin $s$ of the fields at $a_{*} = 0$. Extrapolating this to the 
case $s=0$ he noted that the relation predicted $h(a_{*}=0) = -1.195$.
Our numerical results have shown that in fact, $h(a_{*}=0) = -0.806$, 
accurate to one part in $10^{4}$.

We have studied the evolution of a rotating black hole
emitting massless scalar particles via the Hawking process.
We have concluded that rather than evolve to a 
nonrotating state with $a_{*}=0$, as is often assumed, the hole 
approaches a nonzero value of $a_{*}$. Our numerical results 
have allowed us to ascertain that this value is $a_{*}=0.555$. This 
result confirms the conjecture of Page, and, for the first time,
reveals the exact value of $a_{*}$ to which the black hole 
will relax.  Additional details of this study will be given
in a separate publication, where we shall also report on
additional aspects of black hole-scalar field
interactions. 

%
%

\acknowledgements

CMC is a fellow of The Royal Commission For The Exhibition
Of 1851 and gratefully acknowledges their financial support.
CMC would also like to thank the members of the Relativity
Group at MSU for their continued support. The work of WAH 
and BT was supported in part by NSF Grant No. PHY-9511794. 

%
%

%
%

\end{document}